\documentclass[10pt,prb,aps,twocolumn,floatfix,showpacs]{revtex4-1}
\usepackage{graphicx}
\usepackage{amsmath}
\usepackage{amssymb}
\usepackage{bm}
\usepackage{dcolumn}
\usepackage{subfigure}
\usepackage{psfrag}
\usepackage{float}
\usepackage{subeqnarray}

\begin{document}

\title{Vortex states in nanoscale superconducting squares: the influence of quantum confinement}

\author{L.-F. Zhang}
\email{LingFeng.Zhang@uantwerpen.be}
\author{L. Covaci}
\author{M. V. Milo\v{s}evi\'{c}}
\author{G.R. Berdiyorov}
\author{F.M. Peeters}
\email{Francois.Peeters@uantwerpen.be} \affiliation{Department Fysica, Universiteit Antwerpen, Groenenborgerlaan 171, B-2020 Antwerpen, Belgium}%

\begin{abstract}
Bogoliubov-de Gennes theory is used to investigate the effect of the size of a superconducting square on the vortex states in the quantum confinement regime.  When the superconducting coherence length is comparable to the Fermi wavelength, the shape resonances of the superconducting order parameter have strong influence on the vortex configuration.  Several unconventional vortex states, including asymmetric ones, giant multi-vortex combinations, and states comprising giant antivortex, were found as ground states and their stability was found to be very sensitive on the value of $k_F\xi_0$, the size of the sample $W$, and the magnetic flux $\Phi$.  By increasing the temperature and/or enlarging the size of the sample, quantum confinement is suppressed and the conventional mesoscopic vortex states as predicted by the Ginzburg-Laudau (GL) theory are recovered.  However, contrary to the GL results we found that the states containing symmetry-induced vortex-antivortex pairs are stable over the whole temperature range.  It turns out that the inhomogeneous order parameter induced by quantum confinement favors vortex-antivortex molecules, as well as giant vortices with a rich structure in the vortex core - unattainable in the GL domain.

\end{abstract}


\maketitle
\section{Introduction}

Vortex states in mesoscopic superconductors have been extensively studied in the past two decades, both theoretically and experimentally.\cite{meso1, meso2, meso3, meso4, meso5, meso6, meso7, meso8, meso9, meso20, meso21, meso10, meso11, meso31, gv2, meso32, meso33, meso34, meso35, meso36, meso37, meso38, meso39, meso40, meso41, meso42, meso43}  Two main interactions have been found to govern vortex behavior in a mesoscopic system.  The first one is the vortex-vortex interaction, which causes vortices to form Abrikosov triangular lattices in bulk type-II superconductors.  The second one is the interaction between vortices and sample boundaries, which makes vortex configurations strongly dependent on the size and geometry of mesoscopic samples-whose dimensions are of the order of the penetration depth $\lambda$ or the coherence length $\xi$.  For example, in square mesoscopic samples, vortex configurations try to best match the $C_4$ symmetry.  When there is only one vortex in the sample ($L=1$ state where $L$ is the winding number or vorticity), the vortex always sits in the center of the sample in order to balance the boundary effect from all sides.  For the $L=2$ state, two vortices sit on the diagonal such that the vortex-vortex separation is maximal in order to minimize the intervortex interaction.  A giant vortex with $L=2$ can be induced when the boundary confinement pushes two single vortices together,  as predicted theoretically\cite{meso2} and observed experimentally.\cite{gv1,gv2,gv3}  For $L=3$ state, because of its incompatability with the four-fold symmetry, the theory predicts that the ground state corresponds to an anti-vortex sitting at the center surrounded by four vortices.\cite{anti,anti2}  In short, the symmetry of the sample largely determines the vortex configurations when the size of the superconductor is reduced.

However, the properties of nanoscale superconductors, whose sizes are of the order of the Fermi wavelength $\lambda_F$, are very different from those of mesoscopic superconductors.  This is because the distance between electronic levels becomes comparable to the superconducting energy gap due to quantum confinement \cite{Anderson}.  As a consequence, the number of Cooper pairs is suppressed which leads to the quantum-size effect (QSE),\cite{qse1, qse2, qse3}  quantum-size cascades,\cite{cascades} the shell effect\cite{shell} and inhomogeneous spatial distribution of the order parameter.\cite{qse2}  The latter is the most important for the present work because it is expected to strongly influence the vortex states in nanoscale superconductors. A similar behavior was shown for an isolated vortex core, where oscillations of there order parameter on the order of the Fermi wavelength were predicted. \cite{meso44}

Inhomogeneous superconductivity has been studied in various systems in the last decades and shows more complex behavior than homogeneous ones.  It is known that vortices tend to migrate and get pinned in areas where superconductivity is suppressed.\cite{inrmp}  The reason is that it is more favorable energetically for a vortex to suppress the superconducting order parameter in a region where it has already been suppressed, although sometimes vortices can be pinned where the gap is large.\cite{pin}  Some three-dimensional (3D) samples can also be treated as inhomogeneous systems.\cite{3d1, 3d2, 3Dtip, 3d3, 3d4, 3d5, 3d6}  For example for a 3D tip geometry, an asymmetric $L=1$ vortex state can be the ground state because the thick region prevents the vortex from penetrating it.\cite{3Dtip}  In multi-layered superconductors, vortices enter first and reside favorably in the weak layers.  Then, vortices will penetrate into the strong layers only after weak layers become saturated and various vortex clusters and asymmetric vortex states are induced.\cite{meso6}  Also, the fabrication of anti-dots in superconductors results in a spatially varying superconducting energy gap with a barrier at the interfaces.  In these systems, the combination of the giant vortex, multi-vortex and anti-vortex states can be found as ground state, which depends strongly on the detailed geometry of the antidots. \cite{ando1, ando2,ando3}

For conventional superconductors, $k_F\xi_0 \approx
10^3$ ($k_F$ is the Fermi wave vector and $\xi_0$ is the BCS
coherence length), systems of size comparable to $\lambda_F$ will
not be large enough to host a vortex (being much smaller than the
coherence length). However, materials with small coherence lengths,
e.g. high-T$_c$ cuprate superconductors, will have $k_F\xi_0
\approx 1-4$ and therefore in such systems it is possible to obtain
vortex states in the quantum confinement regime. Another such
system is a graphene flake deposited on top of a
superconductor. Because of the proximity effect, Cooper pairs will diffuse in graphene\cite{gr1, gr2, gr3, gr4}. In graphene the scattering length is large, therefore such a system is in the clean limit. More importantly, near the Dirac point, the Fermi wavelength
is very large and can be easily manipulated by doping. In other
words, $k_F \xi_0$ can be tuned, which will allow for different
vortex patterns to be realized in the graphene flake in the quantum
confinement regime, but for more accessible sample sizes (even above
100nm). A similar configuration was also recently proposed by Knopnin et. al in Ref.[\onlinecite{gr5}]. Yet another
system where effects of quantum confinement on vortex matter can be
probed systematically are the optically trapped cold gases
\cite{xxx}, which are nowadays extremely controllable.


For studying such nanoscale systems, microscopic Bogoliubov-de Gennes (BdG) theory is required.  Previous works used the BdG method to focus on isolated single vortex lines\cite{dos2, dos3, dos4, meso44}, giant vortices\cite{dos5, dos6} and to describe the local density of states modifications due to vortex-vortex and vortex-boundary interactions\cite{dos62, dos7, dos8, dos9} but are in the mesoscopic limit as opposed to the nanoscale limit considered here.  Although Refs.[\onlinecite{meso43}] and [\onlinecite{jp}] studied the groud state vortex states in a mesoscopic-nanoscopic crossover region by solving the BdG equations self-consistently, quantum confinement effects do not play any role.  Recently, we investigated\cite{prl} the vortex states in nano-scale superconducting squares.  We found unconventional vortex states in the quantum limit due to shape-induced resonances in the inhomogeneous Cooper-pair condensate. Vortex-antivortex structures, asymmetric vortex states and vortex clusters were found as ground states over a wide range of parameters.  They are distinct from previous results obtained in mesoscopic superconductors using the GL theory.  However, there are still several aspects that remained unclear.  For example, how does the size of the sample affect the vortex states in nanoscale superconductors?  Under which conditions, does one recover the conventional GL results? Why are the antivortex states more stable in the nanoscale limit while giant vortex states are unfavorable?  How do the vortex states change if temperature is increased?

In order to answer these questions, in this paper, we study  vortex states in nanoscale superconducting squares systematically.  Vortex states for different sample sizes, $k_F\xi_0$ parameters and temperatures $T$ are investigated and the stability of the symmetry-induced vortex/anti-vortex molecules is discussed.  More unconventional states, very different from the ones obtained within GL theory, are found.  This study is therefore fully complementary to what is known for vortex matter in superconductors.

The paper is organized as follows. In Sec.~II, we introduce the theoretical approach, i.e. the BdG approach for a square geometry.  In Sec.~III, we present the inhomogeneous superconducting state in the absence of the magnetic field in order to better understand the QSE in nano-scale superconductors.  In Sec.~IV, the ground states and metastable states are studied at zero temperature and on the sample size dependence of the vortex states is discussed.  In Sec.~V,  the finite temperature ground states are studied.  In Sec.~VI, we discuss the generation of vortex/anti-vortex molecules and study the structure of the vortex core. Finally, we summarize our findings in Sec.~VII.

\section{THEORETICAL FORMALISM}

In the presence of a magnetic field, the Bogoliubov-de Gennes (BdG) equations read
\begin{eqnarray}
\label{BdG}  \begin{pmatrix}
H_e & \Delta(\vec{r}) \\
\Delta(\vec{r})^\ast & -H_e^\ast
\end{pmatrix}
\binom{u_n(\vec{r})}{v_n(\vec{r})}
=E_n\binom{u_n(\vec{r})}{v_n(\vec{r})},
\end{eqnarray}
where $u_n$($v_n$) are electron-(hole-)like quasi-particle eigen-wave functions and $E_n$ are the quasi-particle eigen-energies.  The single-electron Hamiltonian reads $H_e=\frac{1}{2m}(\frac{\hbar\nabla}{i}-\frac{e\vec{A}}{c})^2-E_F$ with $E_F$ being the Fermi energy and $\vec{A}$ the vector potential (we consider a gauge such that $\nabla\cdot\vec{A}=0$). Furthermore, we take the electron band-mass to be isotropic (i.e., $m_x=m_y=m_z=m$) and a circular Fermi surface.  The pair potential is determined self-consistently from the eigen-wave functions and eigen-energies:
\begin{equation}\label{OP}
\Delta(\vec{r})=g\sum\limits_{E_n<E_c}u_n(\vec{r}) v^\ast_n(\vec{r})[1-2f_n],
\end{equation}
where $g$ is the coupling constant, $E_c$ is the cutoff energy, and $f_n=[1+\exp(E_n/k_BT)]^{-1}$ is the Fermi distribution function at temperature $T$.

We consider now a two-dimensional superconducting square with the size $W$.  The confinement imposes Dirichlet boundary conditions (i.e. $u_n(\vec{r})=0,\; v_n(\vec{r})=0,\; \vec{r} \in \partial S$) such that the order parameter vanishes at the surface. In an extreme type-II superconductor (and/or very thin sample), it is reasonable to neglect the contribution of the supercurrent to the total magnetic field.

In this case, the free energy \cite{FreeEg1, FreeEg2} of the system is given by
\begin{eqnarray}\label{FE}
F&=&\sum\limits_{n}\left[2E_nf_n  + k_BT[f_n\ln f_n+(1-f_n)\ln(1-f_n)]\right] \nonumber \\
&+&\int dr \left[-2\sum\limits_{n}E_n|v_n|^2 + 2\Delta(r)\sum\limits_{n}u^{*}_n v_n[1-2f_n] \nonumber \right. \\
&-&\left. g\sum\limits_{n}u^*_nv_n(1-2f_n) \sum\limits_{n'}u_{n'}v^\ast_{n^\prime}(1-2f_{n^\prime}) \right],
\end{eqnarray}
where the spatial dependence of $u_n$ and $v_n$ is implicit.

In order to solve Eqs. (\ref{BdG}) and (\ref{OP}) numerically, we expanded the wave functions $u_n$ and $v_n$ as
\begin{equation}\label{EXPAND}
\binom{u_n(\vec{r})}{v_n(\vec{r})}=\sum_{j_xj_y\epsilon\mathbb{N}^+}\varphi_{jx,jy}(x,y)\binom{u^n_{j_xj_y}}{v^n_{j_xj_y}},
\end{equation}
where the basis set
\begin{equation}\label{BASIS}
\varphi_{jx,jy}(x,y)=\frac{2}{W}sin\left(\pi j_x \frac{x}{W}\right)sin\left(\pi j_y \frac{y}{W}\right)
\end{equation}
is the basis eigen-states of the Hamiltonian $H_e$ in the absence of the magnetic field.  The corresponding eigen-energies of such states are $T_{j_xj_y}= \frac{\hbar^2}{2m} (\frac{\pi}{W})^2 (j_x^2+j_y^2) -E_F$.  Through comparison with results obtained by using the finite difference method, we found that the results are converged when we include the states with energies as large as $\Phi/\Phi_0+5$ times the cutoff energy $E_c$, i.e. $T_{j_xj_y}\in(\Phi/\Phi_0+5)\times [-E_c,E_c]$ where $\Phi_0$ is the flux quantum and $\Phi$ the flux through our sample.

\begin{center}
\begin{table}
  \begin{tabular}{ |  c  |  c  |  c  |  c  |}
   \hline
    Sample & I & II & III  \\ \hline
    $E_F/F_0$       &  4    & 9     & 25   \\ \hline
    $E_c/F_0$       & $30/\pi$   & $30/\pi$ & $50/\pi$  \\ \hline
    $\Delta_0/F_0$  & 1.245   & 1.85 & 3.14 \\ \hline
 $k_F\xi_0$  & 2.04   & 3.09 & 5.07 \\ \hline
  \end{tabular}\\
  \caption{The parameters of the considered sample.  Coupling constant $g=0.4343$ and $E_F$ is measured from the bottom of the quadratic band.}
\label{tab1}
\end{table}
\end{center}

In our numerical investigations, we restrict ourselves to the three materials (or samples) with parameters given in Table \ref{tab1}.  For convenience, we measure the distance in units of the bulk coherence length at zero temperature $\xi_0$ and the energy in units of $F_0=\hbar^2/2m\xi_0^2$ .  Here, $\xi_0={\hbar}v_F/\pi\Delta_0$ where $v_F$ is the Fermi velocity and $\Delta_0$ is the bulk value of the order parameter at zero temperature.  Note that $E_F$ and $\Delta_0$ are not independent when $F_0\varpropto \Delta_0^2/E_F$ up to a constant.

To find the different vortex configurations, which include all stable states, we search for the self-consistent solutions in the following two steps. (1) Global scanning: Starting from any reasonable vortex state (usually, we start from the  Meissner state at $\Phi/\Phi_0=0$), we slowly sweep up/down the applied flux with regular intervals $0.1\Phi_0$ and recalculate the superconducting states each time, until a new state is found.  Then, we repeat the sweeping process from the new state until no new vortex configurations appears.  (2) Special initial states:  Starting from usual states obtained in GL theory\cite{meso10}, we sweep up/down the applied flux.  If a new state appears, we repeat the step (1).  In such a way, we are able to trace back and forth all found vortex states in the whole region of their stability and make sure that the usual GL states are always considered.

\begin{figure}[t]
\includegraphics[width=\columnwidth]{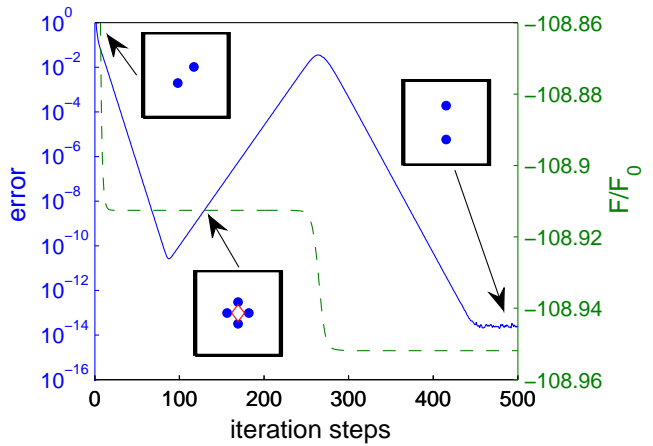}
\caption{(Color online) Error (solid line) and free energy (dashed line) as a function of iteration steps for sample II for $W/\xi_0=5$ at $\Phi/\Phi_0=5$ and $T/T_c=0$.  Three insets show the initial, intermediate and final vortex configurations.  The solid dots and open diamonds in the insets indicate $L=1$ vortex and $L=-2$ giant antivortex, respectively.} \label{tolerance}
\end{figure}

The minimum tolerance in the change of the order parameter  between two steps in the self-consistent iteration is
\begin{equation}\label{Error}
max\left \{ \left | \Delta^{i}(x,y) -\Delta^{i-1}(x,y) \right | \right \}<10^{-13}
\end{equation}
where $\Delta^i$ and $\Delta^{i-1}$ is the order parameter at the $i^{th}$ and $i-1^{th}$ iteration.  We use the absolute tolerance since the relative tolerance can be abnormally high in the vortex core where $|\Delta|\rightarrow 0$.  This is quite strict when we compared to $|\Delta_0|$ but is necessary in order to ensure the precision in finding the true ground states in the BdG calculation.  We show an example the evaluation of the vortex configuration for sample II and $W/\xi_0=5$ at flux $\Phi/\Phi_0=5$.  Based on the GL theory, the ground state of $L=2$ for such a system should be giant vortex state or multi vortex state (for larger squares) where two vortices sit on the diagonal of the sample. We start the calculation with such the multi vortex state as initial state. As seen from the Fig.~\ref{tolerance}, the vortices merge into a vortex anti-vortex molecule and the result converges quickly.  The error between each steps reach as low as $10^{-11}$.  However, the error increases gradually when we continue the self-consistent procedure.  After the second-order phase transition, the new state  with two vortices sitting parallel to one of the sides has lower energy.  Finally, the symmetry of the state does not change and the error is always around $10^{-14}$ which comes from non-physical factors, i.e. numerical accuracy.

In the calculation, we found some situations where results do not converge and this usually comes from the change in the number of the quasi-states contained in the Debye window.  To avoid this, we set the smearing energy $E_S$ for the quasi-states. Then, the self-consistent condition reads
\begin{equation}\label{Smearing}
\Delta(\vec{r})=g\sum\limits_{E_n>0}u_n(\vec{r}) v^\ast_n(\vec{r})[1-2f_n] \times f_n(E_n-E_c),
\end{equation}
where $f_n(E_n-E_c)=[1+\exp(\frac{E_n-E_c}{E_S})]^{-1}$ is the Fermi distribution function.  The choice of $E_S$ is empirical. It should be enough small in order not to affect the results.  Meanwhile, it should be enough large to make the result converge through the iteration.  Our experience show that $E_S/F_0=0.2267$ is suitable for our current work.

%
%
\begin{figure}
\includegraphics[width=\columnwidth]{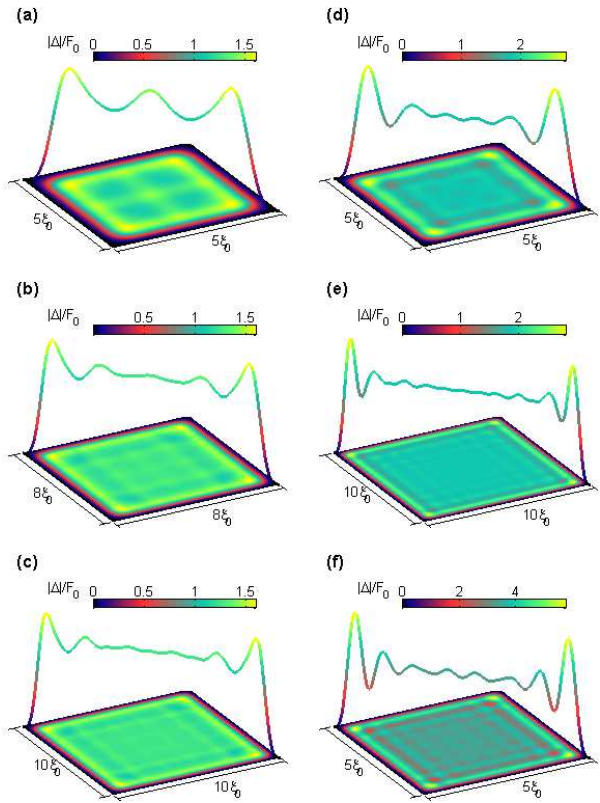}
\caption{(Color online) Contour plots of the order parameter $\Delta(x,y)$ with the corresponding diagonal profiles in the absence of applied magnetic field. Panels (a)-(c) are for sample I with sizes $W/\xi_0=5,8,10$, respectively. Panels (d) and (e) are for sample II with sizes $5\xi_0$ and $10\xi_0$, respectively.  Panel (f) corresponds to sample III with size $5\xi_0$.}
\label{S201}
\end{figure}
%
%
\begin{figure}[h]
\includegraphics[width=\columnwidth]{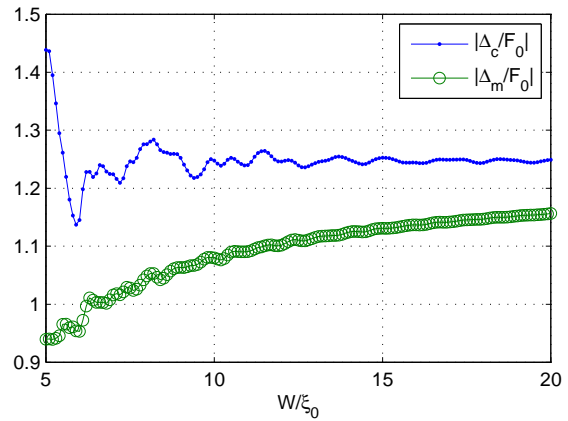}
\caption{(Color online) Size dependence of the order parameter in the center of the sample $|\Delta_c|$ and the spatially averaged $|\Delta_m|$ for sample I.}
\label{S202}
\end{figure}
%
%

\section{Spontaneous Inhomogeneous Superconductivity induced by Quantum confinement}
First, we study the spatial distribution of the order parameter as a function of sample size in the absence of the magnetic field since the vortex configurations will be strongly affected by that distribution once the magnetic field is applied.  In Fig.~\ref{S201} contour plots of the order parameter $\Delta(x,y)$ and the corresponding diagonal profiles are shown for sample I-III with different sizes.  As expected, all the order parameters are fourfold symmetric and show Friedel-like oscillations in space which result in four well defined peaks at each corner.  Hence, the superconductivity is inhomogeneous.  For example, Figs.~\ref{S201} (a)-(c) show the results for sample I with size $W/\xi_0=5, 8, 10$, respectively.  For $W/\xi_0=5$, there are three oscillations in the order parameter along the diagonal and the resonant central peak prevents vortex from sitting here.  However, the profile of the order parameter can be changed dramatically when the size of the sample changes.  For $W/\xi_0=8$ [shown in Figs.~\ref{S201}(b)], we found that the central peak disappears and a relatively flatter area generates in the center.  When $W/\xi_0$ is increased to $10$, the flat area enlarges and the Friedel-like oscillations can be neglected at center when we compared it with oscillations near boundary.  It indicates that the oscillations of the order parameter result from the quantum confinement effect (or boundary effect).

In order to study the quantum size effect on the order parameter, we show the amplitude of the order parameter in the center of the sample $|\Delta_c|$ and the spatially averaged value over the whole sample $|\Delta_m|$ for the sample I with sizes $W/\xi_0=5-20$ in Fig.~\ref{S202}.  As seen, $|\Delta_c|$ changes dramatically with $W/\xi_0$ increasing and converge to $\Delta/F_0=1.245$ when $W/\xi_0>15$.  At the same time, the $|\Delta_m|$ increases gradually with $W/\xi_0$ increasing.  In principle, both of the parameters will converge as $W\rightarrow \infty$ where the boundary effect can be totally neglected.  Since $|\Delta_c|$ and $|\Delta_m|$ show strong quantum size effect between $W/\xi_0=5$ and $10$, we limit ourselves in following sections to study the samples for these particular size .

The profile of $\Delta$ is also strongly affected by $k_F\xi_0$.  Figs.~\ref{S201}(d) and (e) show results for sample II with size $W/\xi_0=5$ and $10$ and (f) is for sample III with size $W/\xi_0=5$.  Comparing to the sample I with the same size, we found that the wave number and the amplitude of the oscillations along diagonal are larger.  It indicates that the superconducting order parameter shows more inhomogeneous behavior with larger $k_F\xi_0$.

%
\begin{figure*}
\includegraphics[width=16cm,height=8cm]{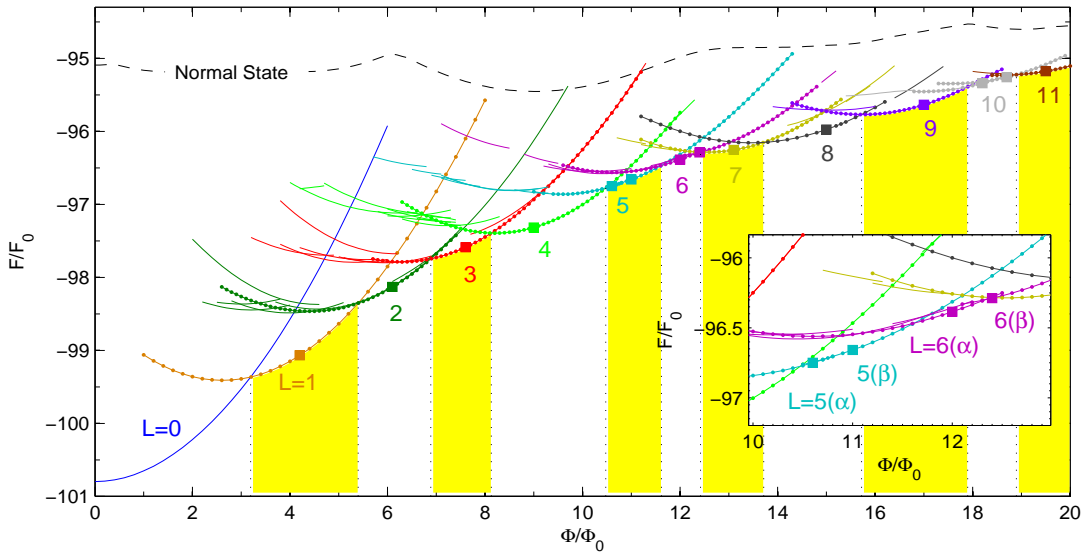}
\caption{(Color online) Free energy as a function of the magnetic flux through sample I with size $10\xi_0\times 10\xi_0$.  Different colors indicate the different winding numbers $L$ and the shaded area indicates the flux range over which the vortex state with winding number $L$ is the ground state.  The vortex configurations of the ground state marked by solid squares are shown separately in Fig.~\ref{VCg} and the corresponding free energy curves are marked by dots.  The inset zooms in the region where $L=5$ and $6$ are the ground states} \label{FHall}
\end{figure*}
%

\section{Vortex configurations at zero temperature}

In this section, we consider the zero temperature case for which the system is always in the quantum limit ($T<1/k_F\xi_0$).
First, we study the sample I with size $10\xi_0 \times 10\xi_0$ and show in Fig.~\ref{FHall} the free energy of the stable vortex states for flux $\Phi/\Phi_0 \in [0,20]$. Different curves (colors) represent states with different winding number $L$ and the states among them which reached the ground state are marked by dots.  Vertical lines and shadows show the flux range for each $L$ state as the ground state.  The top dashed line stands for the free energy of the system in the normal state when the coupling constant $g$ is set to zero.  When compared with the GL theory\cite{meso10}, one of the differences is that the free energy of the normal state depends on the magnetic field while it is a constant in GL theory.  The reason is that the energy levels of the confined electrons are different for different magnetic fields.  In our case, the change of the energy is relatively small when compared with the energy gap (energy difference between the normal state and the superconducting state) especially in weak fields.  Although the shown energy curves look conventional, there are significant differences with the GL theory.\cite{meso10}

\begin{figure}[t]
\includegraphics[width=\columnwidth]{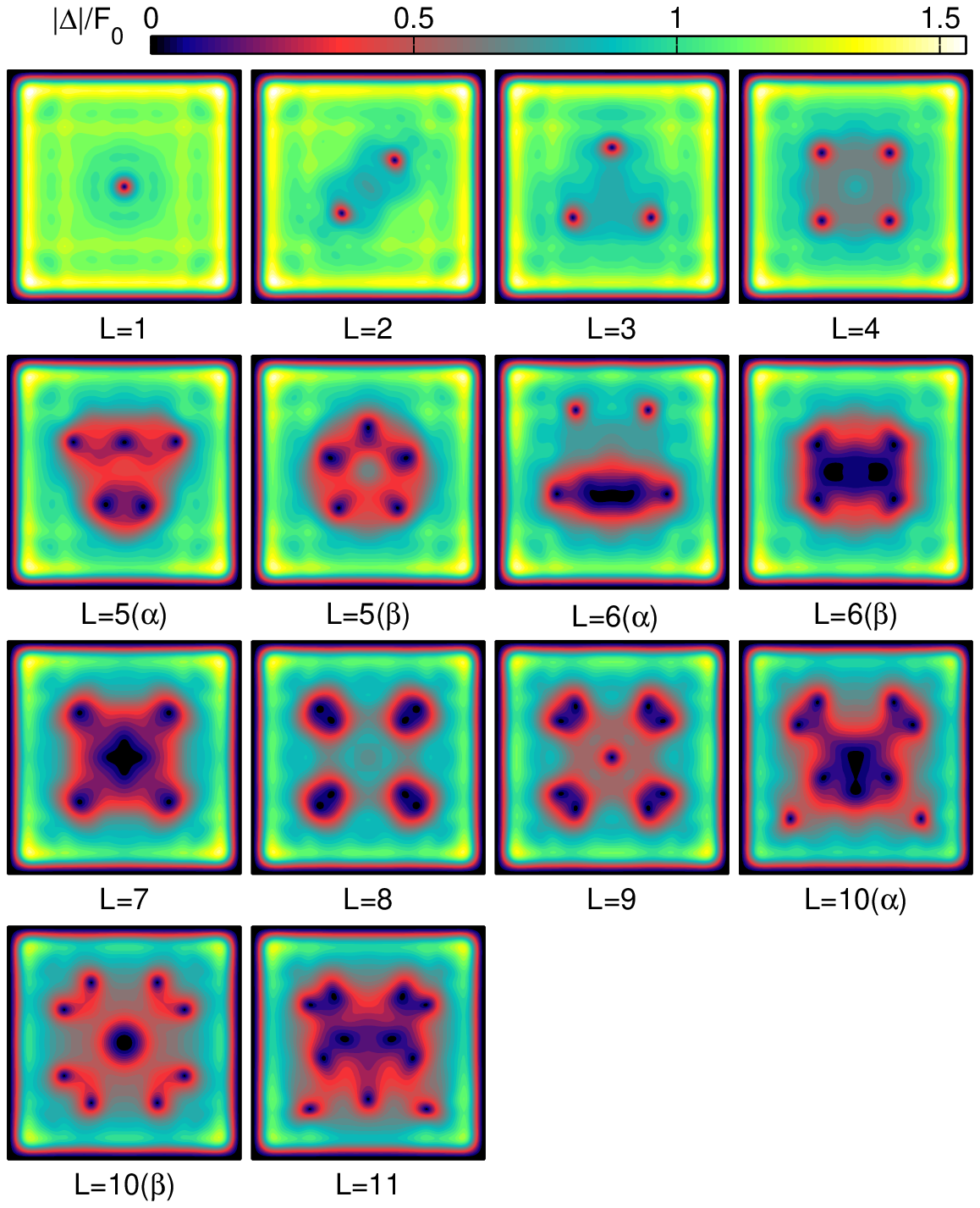}
\caption{(Color online) Contour plot of the amplitude of the order parameter for the ground states marked by solid squares shown in Fig.~\ref{FHall}. $L=10(\alpha)$ and $(\beta)$ are for flux $\Phi/\Phi_0=18.2$ and $18.7$, respectively.  Darkest areas indicate the positions of vortices.  Note that at the center an anti-vortex (for $L=7$) or a giant vortex [for $L=10(\beta)$] can spontaneously form. Note also that for some vorticities two different ground states are found (at different magnetic fields).} \label{VCg}
\end{figure}

By sweeping the magnetic field up and down, we can get the full energy spectrum and the corresponding vortex states.  For a certain magnetic field, it is common to have more than one converged solution.  The lowest energy state is the ground state while the states with higher energy are referred to as metastable states.  In Fig.~\ref{VCg}, we show the contour plot of the absolute value of the order parameter of the corresponding ground states for various winding numbers.

As can be seen from Fig.~\ref{FHall}, the system favors states with winding numbers $L=1,4,8$ and $9$ because they have relatively large ground state flux range (excluding the Meissner state).  From Fig.~\ref{VCg}, we observe that these states have fourfold symmetry which is compatible with the sample geometry. One interesting feature of this system is the richness of metastable states.  These states appear for all winding numbers $L$ except $L=0$ and $1$.  The number of metastable states reaches a peak for $L=4$ and equals $11$.  From the free energy curves, we notice that the energy difference between the ground state and the metastable states can sometimes be very small.  This makes the ground state difficult to find in simulations unless we sweep the field up and down many times.  We also note that most metastable states are focused at lower magnetic fields from the corresponding ground state flux range.  The reason is that vortices get easily stuck at the boundary due to the pronounced oscillations of the order parameter.  For the same reason, their stability range is narrower due to asymmetry.

The number of metastable states decreases for higher $L$.  In this case, the shorter distances between vortices cause strong interaction between them.  This limits the choice of stable positions for vortices and therefore the number of metastable states are lower.  Due to this reason, metastable states are less favorable for smaller samples because of easy saturation with vortices.  For example, no metastable states were found in sample I with size $W/\xi_0=5$.

%
%
\subsection{Ground states}
Next we discuss the ground states configurations for different vorticities in more detail.  For the $L=1$ state, shown in Fig.~\ref{VCg}, the vortex sits at the center of the sample which is compatible with the conventional picture.  Although such result is observed for different parameters of the sample, the state with diagonal location of the vortex can also be found in some cases.\cite{prl}  It is clear that the order parameter around the vortex core is suppressed and the profile shows the competition between $C_4$ symmetry and $C_\infty$.  It means that the vortex has long range (longer than $\xi_0$) interaction with other objects such as the other vortices and/or boundaries.

By increasing the flux to $\Phi/\Phi_0=5.24$, the ground state shifts from $L=1$ to $L=2$ and two vortices sit along the diagonal.  This again coincides with the result from GL theory, and results from the competition between the confinement imposed on vortices by the Meissner currents and the vortex-vortex repulsion.  In this case, these effects can be clearly seen from the profile of the order parameter, especially from the suppressed area around the vortices.  The vortex-vortex interaction suppresses the order parameter mostly in the area between them.  This can not be found in GL theory because the order parameter is always smooth and changes slowly in space.  The state with $L=3$ shown in Fig.~\ref{VCg} becomes the ground state when the fieldapplied flux is between $\Phi/\Phi_0=6.9$ and $8.12$.  It resembles the multi-vortex state obtained within GL theory where the three vortices are at the apices of a equilateral triangle.  However, the perpendicular bisector of the triangle always coincides with one of the diagonals of the square sample in GL theory while it is parallel to one of the edges in our case.  This is because the grid-like pattern in the inhomogeneity of the order parameter imposes preferential positions for the vortices inside the square.  The state $L=4$ has a similar feature but the configuration is compatible with the GL result.

From these states, we conclude that when the GL vortex configuration, which minimizes the vortex-vortex and vortex-boundary interaction, matches the oscillation pattern due to quantum confinement, then the state has a wider flux stability range.

Two ground states, $L=5(\alpha)$ and $L=5(\beta)$, are found for $L=5$ in the flux ranges $\Phi/\Phi_0= 10.48-10.78$ and $10.78-11.62$, respectively.  Both of them have a pentagonal vortex configuration.  This is because the particular shape resonance at the considered field causes the order parameter to be peaked at the center. Therefore, it costs energy for vortices to sit in the center of the sample.  For the same reason, the ground states $L=6(\alpha)$ and $L=6(\beta)$ do not have vortices in the center.  Moreover, when $L\geqslant 6$, vortices start to be compressed in the sample.  If they do not form giant vortices, they will be very close to each other and form string-like structures [see $L=6(\alpha)$ state].

States with $L=7$, $L=8$, $L=9$ and $L=10(\beta)$ shown in the panels of Fig.~\ref{VCg} have a common feature, as all of them keep the fourfold symmetry.  $L=7$ contains a antivortex at the center while $L=9$ and $L=10(\beta)$ have a single vortex and a giant vortex with $2\Phi_0$ at the center, respectively.  The state with $L=7$ is the only ground state which contains an antivortex.  The antivortex is closely surrounded by four vortices and forms the core structure for $L=7$.  The outer shell is formed by the remaining four individual vortices sitting at four corners.  States with $L=8$ and $L=9$ contain vortex dimers, i.e. two vortices close to each other at each corner.  The fourfold symmetry makes both former states have a larger ground state flux range.  $L=10(\beta)$ also keeps the $C_4$ symmetry but the energy is sometimes even higher than the state $L=10(\alpha)$, which has only $C_2$ symmetry.  The reason is that the giant vortex costs extra energy.

\begin{figure}[p]
\includegraphics[width=\columnwidth]{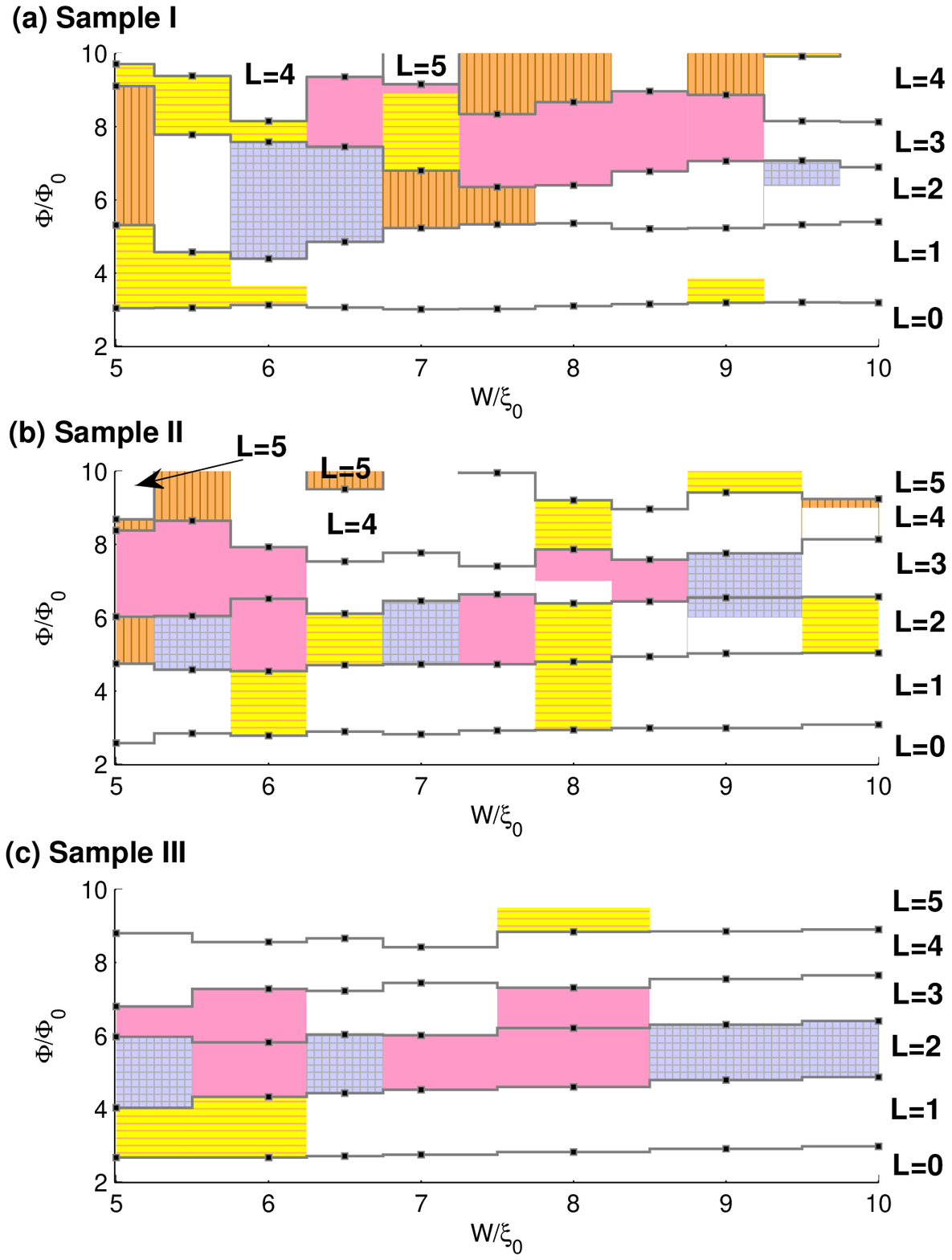}
\includegraphics[width=\columnwidth]{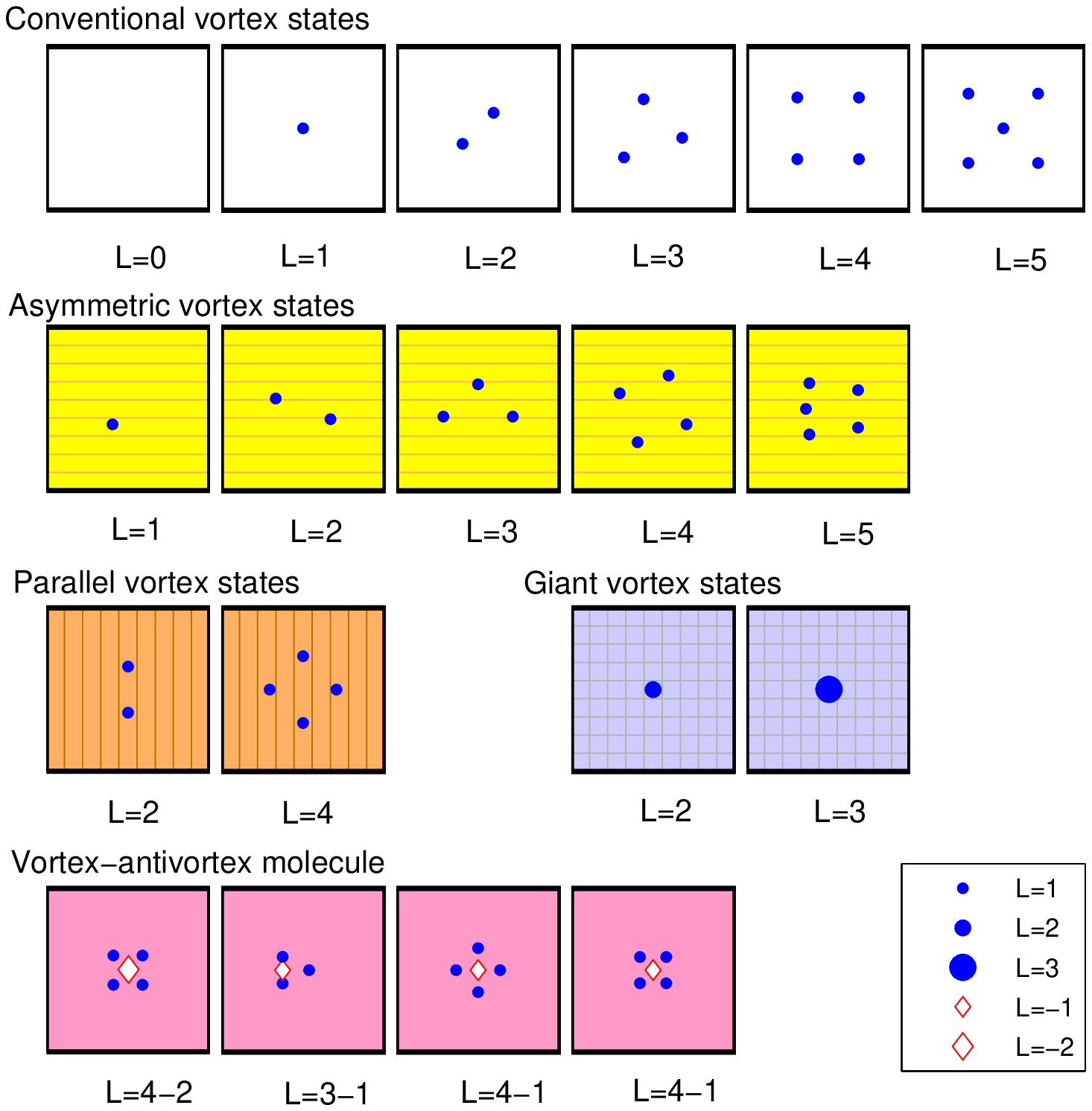}
\caption{(Color online) Phase diagram for samples I-III.  Different shadings of the background indicate different types of vortex states.  Black(blue) dots indicate vortices, and diamonds indicate antivortices in the schematic diagrams of vortex configurations (bottom figures).  The symbols are larger when vortices contain multiple flux quanta. } \label{Phase}
\end{figure}

In order to visualize the changes in the ground states when key parameters change, we plot the phase diagram for samples I-III for $W/\xi_0=5-10$ and $\Phi/\Phi_0=0-10$ in Fig.~\ref{Phase}.  Different shadings of blocks in Fig.~\ref{Phase} indicate different vortex types.  The plain white background represents conventional multi-vortex states as found within the GL theory, while the blue background with square grid represents giant vortices, also compatible with the result obtained from GL theory.  Asymmetric vortex states attained only by BdG theory are represented by yellow background with horizontal grid pattern. States containing parallel vortex chains, represented by orange background with vertical grid pattern, and part of the vortex-antivortex molecules represented by pink(grey) background, are new compared to GL theory.

%
%

We can conclude from Fig.~\ref{Phase} that the quantum size effect is important not only for the transition field and the stability range of the different vortex states, but also for the vortex configurations.  The reason is that the oscillation patterns of the order parameter are very sensitive to $k_F\xi_0$ and may cause totally different behaviors even for two samples of identical size.

As can be seen from Fig.~\ref{Phase}(a) and (b), the vortex phase transition fields vary greatly with sample size $W$ except for the phase boundary between $L=0$ and $L=1$ state.  All the phase boundaries oscillate with $W$.  We find that some samples favor vortex states with even winding number $L$ while other disfavor them.  For example, sample I with $W/\xi_0=5$ favors $L=2$ state whereas the one with $W/\xi_0=7$ disfavors $L=4$ to the point of non-existance.  When $W$ becomes large, the transition fields start to converge and the flux stability range of each $L$ state will be roughly one flux quantum.

When compared to samples I and II, sample III (with large $k_F\xi_0$) shows a more conventional picture.  Moreover, the phase transition field increases only slightly with increasing $W$.  It means that for large $k_F\xi_0$, the quantum size effect on the transition field, at least when winding number $L$ is small, can be neglected. Nevertheless, a plethora of different vortex configurations is found.  For example, an asymmetric $L=1$ state is found in all three samples.  Moreover, sample I always shows asymmetric states when $W/\xi_0<6$. One other interesting phenomenon is that, for $L=2$, sample II can host all five types of vortex states with $W$ increasing.

From the phase diagram in Fig.~\ref{Phase}, we notice that nanoscale superconductors favor anti-vortices and disfavor giant vortices.  For example, the giant vortex state appears for $L=2$ in sample II with $W/\xi_0=7$.  Based on GL theory, only smaller samples will exhibit a giant vortex configuration.  However, when $W/\xi_0<7$, the two individual vortices form an asymmetric vortex state. On the other hand, we find that the probability of forming anti-vortices is much higher than in GL theory.  In GL case, antivortex states usually appear for $L=3$ when four vortices are at the four corners and surround a centered antivortex.  Usually the distance between vortices and the antivortex is small (less than $\xi$).  In the BdG calculation, at least two more antivortex states can be found.  One is still the $L=3$ state and the antivortex is still at the center but four vortices are at the edges instead of the corners.  This configuration was first presented by us in Ref.~[\onlinecite{prl}] where we showed that the size of the vortex/anti-vortex(V-aV) molecule is larger than the one obtained with GL theory.  The other antivortex state appears for $L=2$.  In this case, the four vortices still sit at four corners but the centered antivortex carries two flux quanta, e.g. it is a giant antivortex!  Due to the strong vortex-antivortex interaction, the size of such V-aV molecule is small.

%
%
\begin{figure}[t]
\includegraphics[width=\columnwidth]{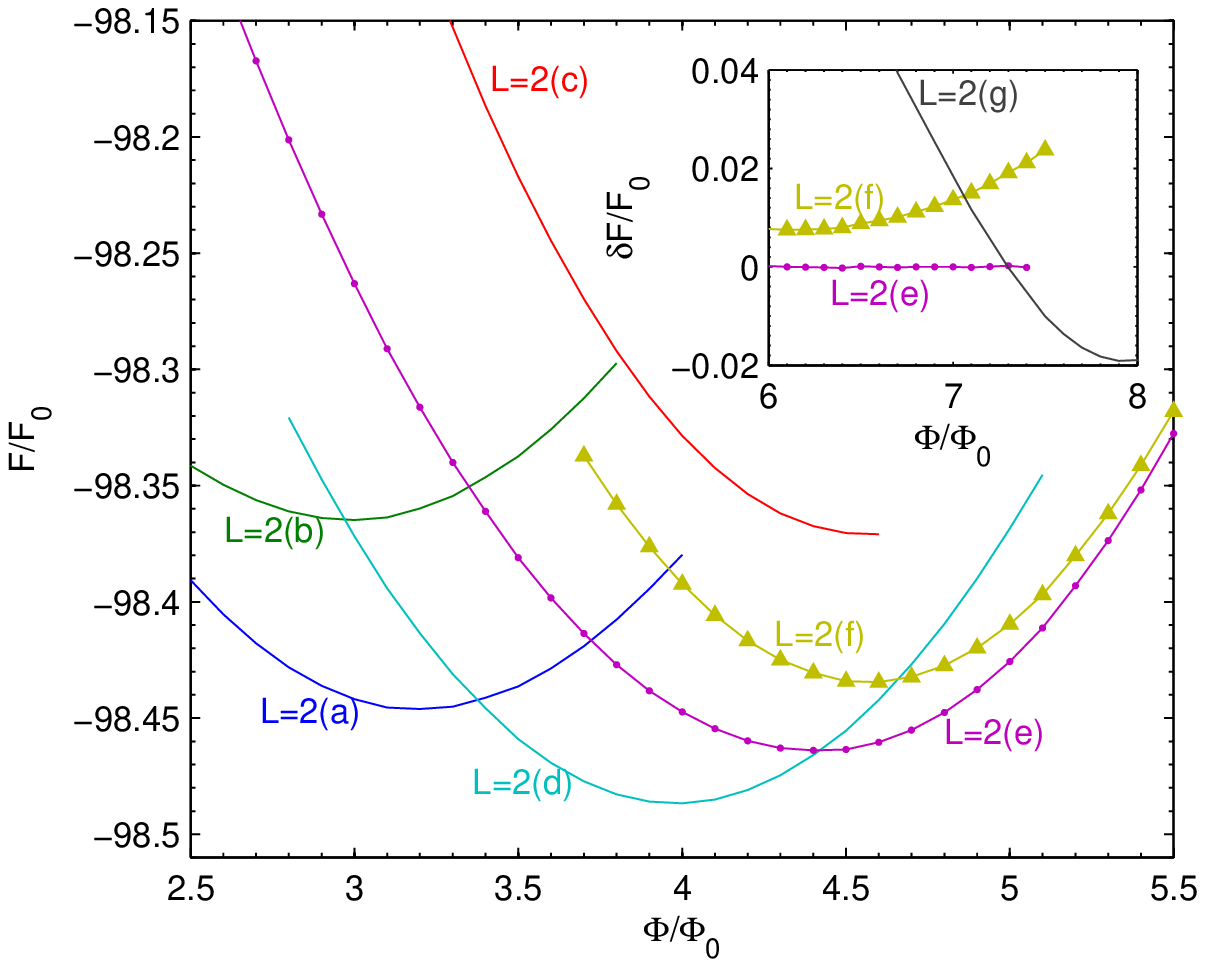}
\includegraphics[width=\columnwidth]{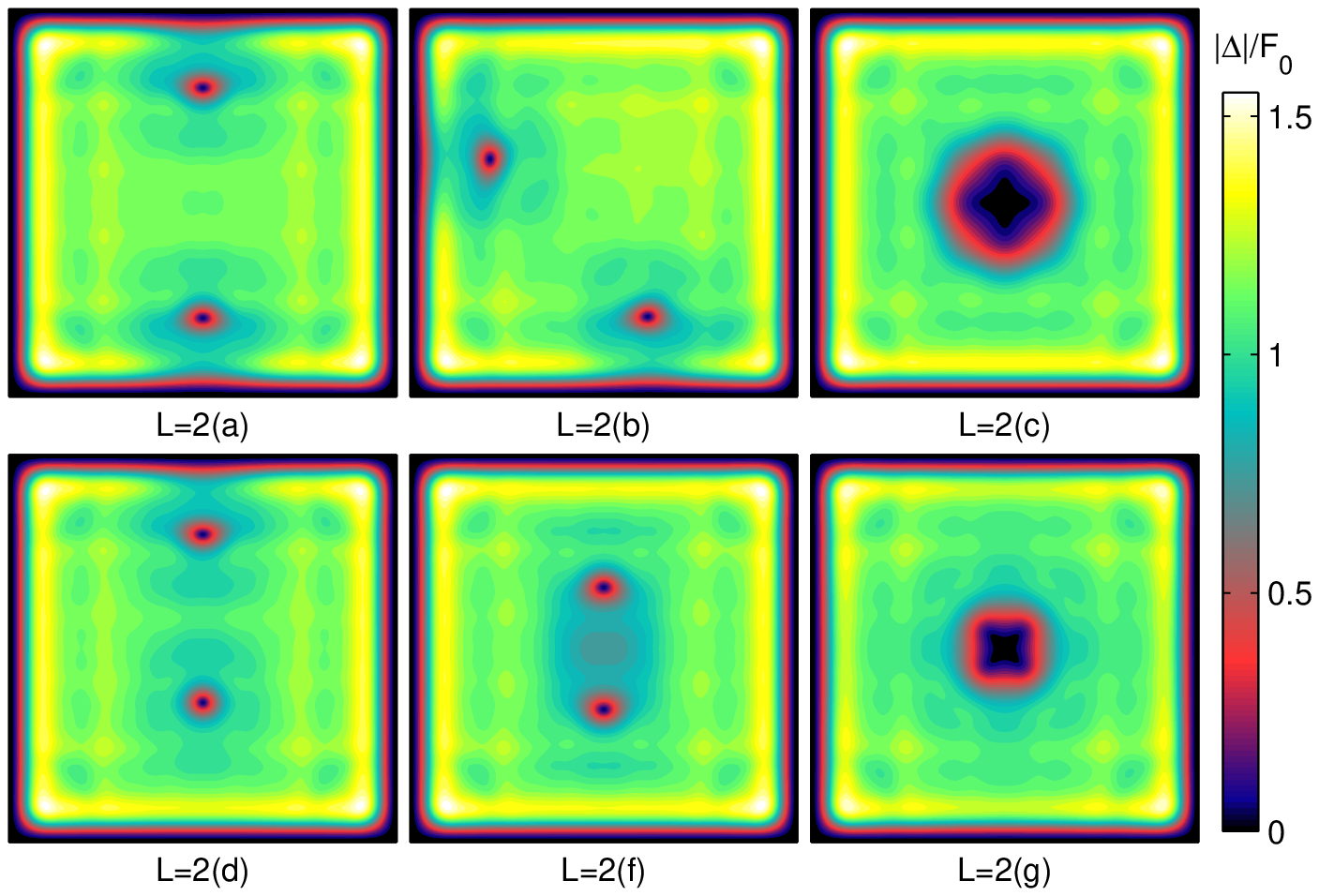}
\caption{(Color online) Free energy curves of states with $L=2$ in sample I with size $W/\xi_0=10$ (top panel) and the corresponding contour plots of the order parameter (bottom panel).  Note that the contour plot of the ground state $L=2(e)$ is shown in Fig.~\ref{VCg}.  The inset shows the energy difference between the ground state $L=2(e)$ and the metastable states $L=2(f)$ and $L=2(g)$ at higher field.} \label{FHL2}
\end{figure}
%
%
\subsection{Metastable states}
Next we will briefly discuss the metastable states of this system.  To do this, we start from sample I with $W/\xi_0=10$.  Metastable states are important in the BdG formalism because the energy difference between the ground and the metastable states can be very small. This suggests that these states could be easily found in experiments.  Alternatively, some metastable states can become ground states as the parameters are changed.

All six found metastable states for $L=2$ and their free energy curves are shown in Fig.~\ref{FHL2}.  The state $L=2(f)$ is similar to the ground state $L=2(e)$, but rotated over $45^{\circ}$, hence their free energies are very close to each other.  Actually, the difference in the orientation of the vortex pattern always results in a small difference in energy.  State (f) is not obtained within the GL theory.  In our case, due to the shape-resonant inhomogeneity of the order parameter, the rotation of the vortex pattern to the ground-state configuration is prevented by the spatial oscillations of the order parameter.

The metastable state $L=2(g)$ is only stable at higher field and its free energy is very close to the ground state $L=2(e)$.  Therefore we zoomed on the energy difference in the inset of Fig.~\ref{FHL2}.  From the figure, one sees that the energy of the $L=2(g)$ state is lower than the ground state, $L=2(e)$, when the applied flux is larger than $\Phi/\Phi_0=7.3$.  In fact, the state can exist even up to $\Phi/\Phi_0=10$.  From the vortex configuration shown in Fig.~\ref{FHL2}, we find that this state is a giant vortex.  Such a state has been predicted by the GL theory because the magnetic field pushes the two vortices towards each other and makes them merge into a giant vortex.  Usually, the phase transition between the multi vortex state and the giant vortex state is continuous (second-order).  However, the barrier induced by the inhomogeneity of the order parameter leads to a first order phase transition in our case.  One more difference between the BdG giant vortex and the one in GL theory is its core structure.  Due to the shape resonances, the contour plot of the core shows a diagonal cross shape while the giant vortex core in the GL case is always circular.  Furthermore, the giant vortex state in our results has two allotropes: see state $L=2(c)$, compared to the state $L=2(g)$.  The $L=2(g)$ state exists up to higher field while $L=2(c)$ only exists in lower field.  Hence, the size of the giant vortex seen in $L=2(c)$ is larger than the one seen in $L=2(g)$.  Another difference between them is the orientation of the core.  $L=2(g)$ has diagonal cross shape while the state $L=2(c)$ has edge cross shape.

\begin{figure}[t]
\includegraphics[width=\columnwidth]{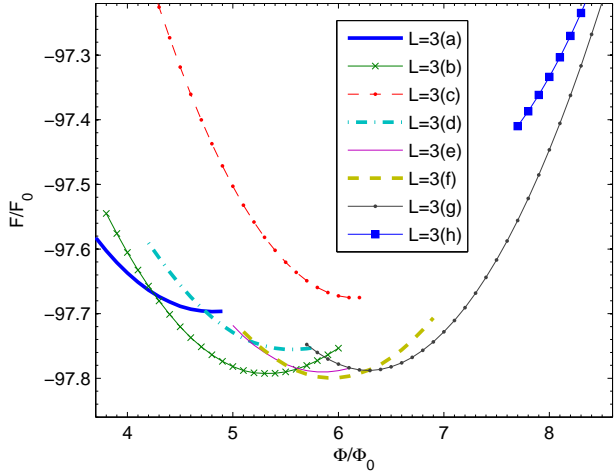}
\includegraphics[width=\columnwidth]{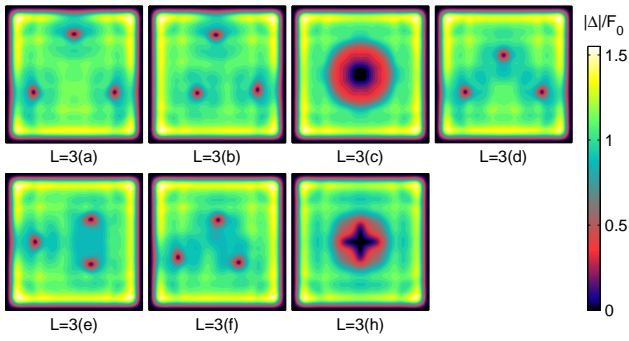}
\caption{(Color online)  Free energy curves of states with $L=3$ in sample I with size $W/\xi_0=10$ and the corresponding contour plots of the order parameter. The contour plot of the ground state $L=3(g)$ is shown in Fig.~\ref{VCg}.} \label{FHL3}
\end{figure}
%

Other three metastable states $L=2(a)$, $L=2(b)$ and $L=2(d)$ are observed only in lower field.  They have in common the fact that at least one vortex is stuck at the boundary since the Meissner current pushes the vortex outward at low fields.  It is obvious that they have lower energy when the applied field is lower. The energy of state $L=2(a)$ is always lower than the one of state $L=2(b)$ because of the longer distance between the vortices.  As is usual in mesoscopic superconductors, vortices in these states avoid to be located at the very corners of the sample, due to strong local superconductivity there.

For the $L=3$ metastable states, the results are summarized in Fig.~\ref{FHL3}.  The state $L=3(h)$ is a V-aV state and exists only in higher field while the giant vortex $L=3(c)$ only exists in lower field.  Note again that the energy of the giant vortex states is much higher than the other metastable states with three single vortices.  $L=3(f)$ has lowest energy for $L=3$ around $\Phi/\Phi_0=6$, when there is one vortex located at the boundary. States $L=3(b)$ and $L=3(a)$ follow when the field  decreases and there are two and three vortices stuck at the boundaries, respectively.  States $L=3(d)$ and $L=3(e)$ are disfavored and have higher energy due to the close distance between vortices.  Note again that no vortex sits at the corners in this states.

%
\begin{figure}[t]
\includegraphics[width=\columnwidth]{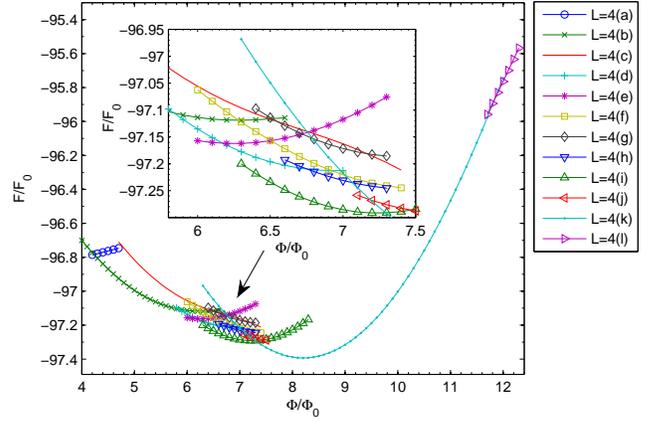}
\includegraphics[width=\columnwidth]{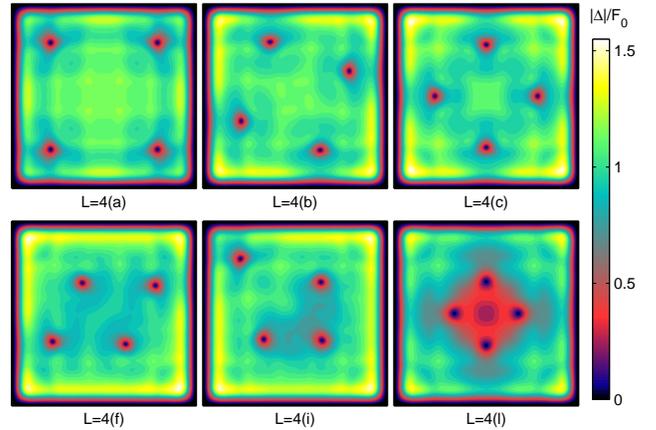}
\caption{(Color online) Free energy curves of vortex states with $L=4$ in sample I with size $W/\xi_0=10$ and the contour plots of the order parameter of selected vortex states.  Inset shows details of the free energy curve around $\Phi/\Phi_0=6.5$.  The contour plot of ground state $L=4(k)$ is shown in Fig.~\ref{VCg}.} \label{FHL4}
\end{figure}
%
%

States with $L=4$ show a wide ground state flux range and $11$ different metastable states, which is the largest variety of all $L$ vortex states.  From Fig.~\ref{FHL4}, we find that the metastable states concentrate around the applied flux $\Phi/\Phi_0=6.5$.

For low fields, we conclude again that vortices are close to the surface and these states are always the lowest energy state for a given $L$ state at low fields.  States $L=4(b)$ and $L=4(c)$ have $C_4$ symmetry and all of the four vortices are trapped close to the boundary.  Please note that in the state $L=4(a)$ four vortices sit at the corners.  This kind of state is rare in the BdG results because corners give the highest potential energy contribution for vortices.  From the free energy curve of the state $L=4(a)$, we can find that the slope of the energy curve is opposite to the other $L=4$ states [such as $L=4(b)$] in this field range.  This indicates that vortices are repelled by the Meissner current in order to balance the inward force the vortices experience from the corner.  When the field is too low, a vortex is expelled from the sample and the state jumps to a $L=3$ state.  The vortex configuration $L=4(f)$ has been found experimentally in conventional mesoscopic superconductors\cite{meso9} but was a result of the presence of pinning sites.  This state can not be obtained in plane squares within the GL theory .  Another state to notice is $L=4(i)$ whose energy curve does not cross any other $L=4$ curve.  When the superconductor is in this state and the field is swept down, this state will be the first to jump to the $L=3$ state.  This is understandable from the vortex configuration of $L=4(i)$ because the vortex at the corner is easily expelled when field is lowered.

At high fields only one metastable state exists $L=4(l)$. It can be seen as the state obtained after a $90$ degrees rotation of the ground state-$L=4(k)$.  This is a consequence of the fact that the inhomogeneous pattern of the order parameter changes with field.  At such a high field, the corner vortex position in state $L=4(k)$ becomes unstable, which forces the vortices to sit at the edges, similar to the $L=4(l)$ case.  At the same time, the strong field pushes vortices closer together so that the distance between vortices in $L=4(l)$ is shorter than the one in the ground state $L=4(k)$.

\section{Vortex states at finite temperature}

\begin{figure}[ttt]
\includegraphics[width=\columnwidth]{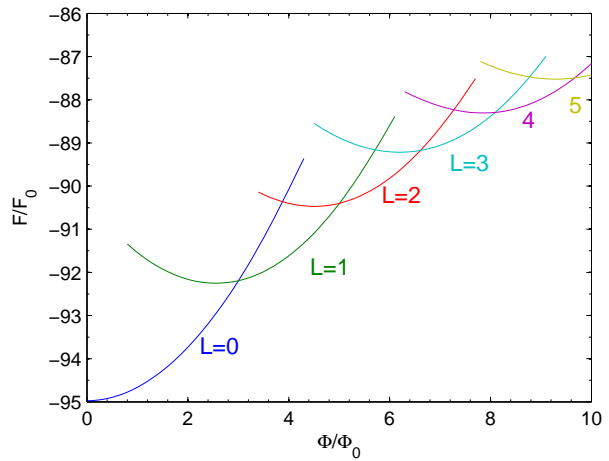}
\caption{(Color online) Free energy as a function of the magnetic flux through the square sample I for size $W/\xi_0=10$ and $T/T_c=0.6$.} \label{HF_T2}
\end{figure}

So far, all our calculations were done at zero temperature, $T/T_c=0$, where $T_c$ is the bulk critical temperature at zero flux $\Phi/\Phi_0=0$.  In what follows, we investigate the effect of temperature on the vortex configuration.  First, we show all the vortex states for the flux range $\Phi/\Phi_0\in[0,10]$, for sample I with size $W/\xi_0=10$ at $T/T_c=0.6$ where the system is NOT in the quantum limit since $T/T_c>1/k_F\xi_0$.  The corresponding free energy curve as a function of flux is presented in Fig.~\ref{HF_T2}.  Contrary to the results for $T/T_c=0$, which were shown in Fig.~\ref{FHall}, the figure looks more conventional (similar to the results obtained by GL theory in Ref.~\onlinecite{meso10}) and there is only one stable state for each winding number $L$.  Moreover, only giant vortex states are found in this case for $L\geqslant 2$.  For the size of the square sample considered here, $W \approx 10\xi$, GL theory\cite{meso10} predicts that multi-vortex states should exist.  Here we find instead that multi-vortex states are absent since $\xi$ increases as temperature increases.

\begin{figure}[ttt]
\includegraphics[width=\columnwidth]{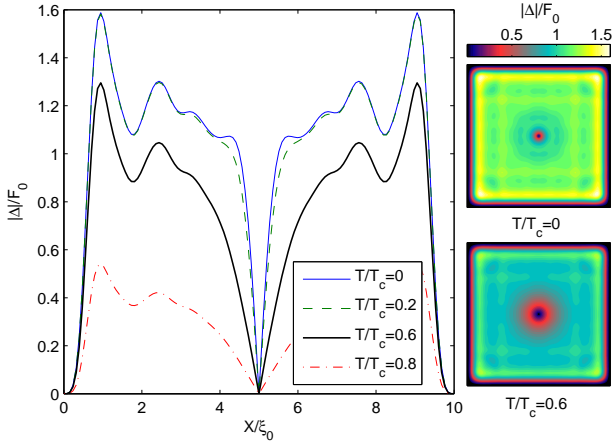}
\caption{(Color online) The diagonal profile of the order parameter for sample I with size $W/\xi_0=10$ for $\Phi/\Phi_0=4$ and $L=1$ at different temperatures.  The corresponding contour plots of the order parameter at $T/T_c=0$ and $T/T_c=0.6$ are also shown.} \label{xiD}
\end{figure}

In order to see how temperature affects the coherence and the profile of the superconducting order parameter, we show the order parameter for $\Phi/\Phi_0=4$ and $L=1$ in Fig.~\ref{xiD}.  The diagonal profile of $|\Delta|$ at $T/T_c=0$ shows the strongest Friedel-like oscillations.  As temperature increases to $T/T_c=0.2$, the profile is similar to the one obtained at zero temperature, but with less oscillations at the vortex core.  Both cases are in the quantum limit and the order parameter shows rapid variation in the core.  When temperature reaches $T/T_c=0.6$, we find that both the average and the oscillations of the absolute value of the order parameter are suppressed which indicates that the vortex states become more conventional.  Finally, the order parameter is smooth at $T/T_c=0.8$ and the GL results are approached.

\begin{figure}[ttt]
\includegraphics[width=\columnwidth]{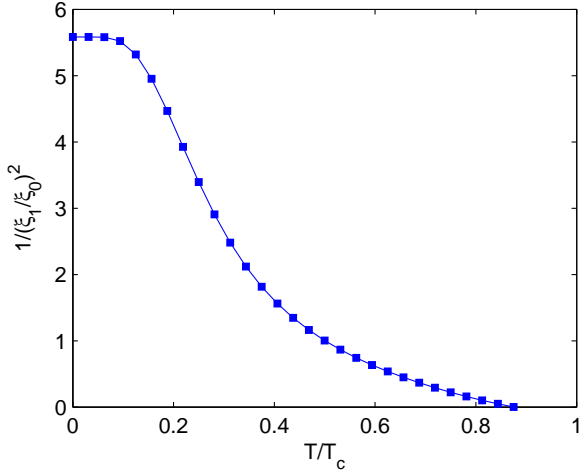}
\caption{(Color online) Temperature dependence of $(\xi_0/\xi_1)^2$ for sample I with size $W/\xi_0=10$ and $\Phi/\Phi_0=4$.} \label{xiT}
\end{figure}

As seen from the Fig.~\ref{xiD}, the coherence length, which represents the vortex core radius, increases with increasing temperature.  As defined by Kramer and Pesch\cite{kramer}, we calculate the coherence length $\xi_1$ as
\begin{equation}\label{xi}
\frac{1}{\xi_1}=\lim_{r \to 0}\frac{\Delta(r)}{r\Delta_0}
\end{equation}
where $r$ is the distance to the vortex core.  We plot in Fig.~\ref{xiT} $(\xi_0/\xi_1)^{2}$ as a function of temperature, $T/T_c$ .  As discussed in Ref.~[\onlinecite{dos3}], $\xi_1$ can be described by $\xi(T)\propto (T_c-T)^{-1/2}$ when $T$ is close to $T_c$ ($T/T_c>0.5$ in our case).  In the intermediate temperature regime, there is a substantial suppression of the coherence length because of the bound states.  At low temperature, the shrinkage of the coherence length stops and saturates when the system is in the quantum limit.  Note that $\xi_1$ at $T/T_c=0.6$ is around three times larger than the one at zero temperature.  This explains why only giant vortex states can be found at such temperatures.

\begin{figure}[t]
\includegraphics[width=\columnwidth]{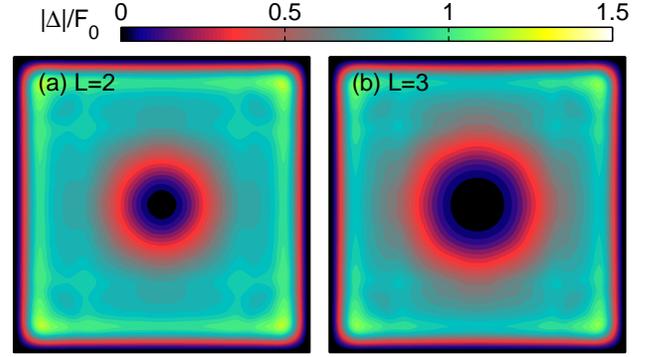}
\caption{(Color online) The contour plots of the order parameter for sample I with size $W/\xi_0=10$ at $T/T_c=0.6$ for (a) $L=2$ and (b) $L=3$.} \label{DeltaT2}
\end{figure}

Fig.~\ref{DeltaT2} shows the order parameter for sample I at $T/T_c=0.6$ for $L=2$ in panel (a) and $L=3$ in panel (b), respectively.  Both are giant vortex states and the $C_4$ symmetry grid pattern is strongly suppressed.  As can be seen from the figure, the vortex cores show perfect circular symmetry, which is in agreement with the results from GL theory.  Of course, the size of the vortex core shown in panel (b) is larger than the one shown in panel (a) because its vorticity is larger.

%
%

\begin{figure}[b]
\includegraphics[width=\columnwidth]{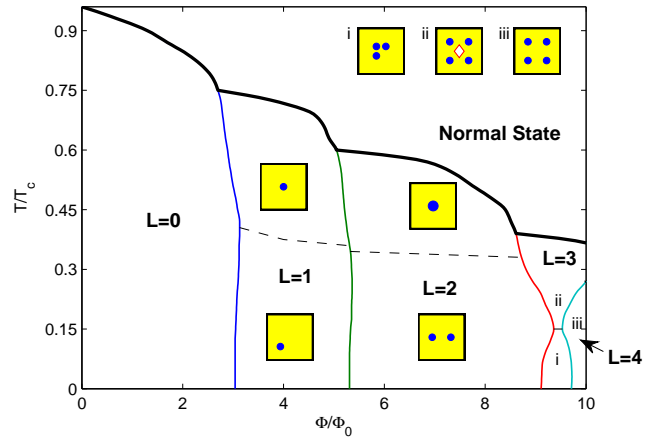}
\caption{(Color online) Temperature-Flux phase diagram for sample I with size $W/\xi_0=5$.  The vortex configurations of areas (i-iii) are shown as insets in the upper right corner.} \label{HTphase}
\end{figure}

Finally, we end this section with the (T-$\Phi$) phase diagram for lower fields for sample I with $W/\xi_0=5$. This is shown in Fig.~\ref{HTphase}. The thick black curve indicates the phase boundary between the superconducting and the normal state.  When the system is in the quantum limit, for these parameters, only unconventional vortex states, such as asymmetric $L=1$ and $L=3$ states and edge-parallel $L=2$ states, are found as ground states.  When temperature increases, the vortex states become conventional and the $C_4$ symmetry of the states is always preserved.  Note that the asymmetric $L=1$ state goes through a continuous phase transition to the symmetric $L=1$ state, which means the vortex moves gradually as the temperature changes. However, for higher winding numbers, the system usually goes through a first order phase transition.  For example, the phase transition between the parallel vortex state and the giant vortex state of $L=2$ is of first order.  This is different from the GL result, where vortices merge into a giant vortex through a continuous phase transition\cite{benxu}.  For the $L=3$ state, we note that the ground state flux range for the four-fold symmetric V-AV state is larger than the asymmetric one due to the compatibility of its symmetry with the geometry of the sample.

Concluding this section, higher temperature: 1) makes vortex states look more conventional (closer to the GL results); 2) smoothens the order parameter; 3) suppresses the influence of the oscillation of the order parameter and 4) increases the superconducting coherence length $\xi$.  As a consequence, the number of metastable states is also lowered.  The effect of temperature is very different (more complex) from the effect obtained by simply changing the effective size of the sample as is usually done within the GL theory.

%

\section{Giant anti-vortex and the structure of the vortex core}
%
\begin{figure}[t]
\includegraphics[width=\columnwidth]{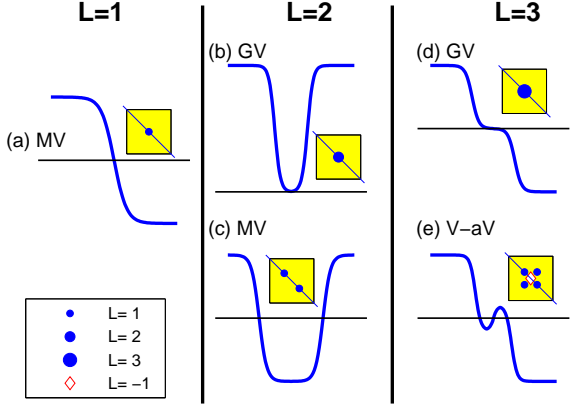}
\caption{(Color online) Diagonal profile of the order parameter for different vortex states in the GL theory.   Blue thick curves represent the order parameter.  All the phases have been adjusted such that on the diagonal the order parameter is real.  The horizontal line indicates the zero of the order parameter and a vortex appears when it intersects the order parameter.  Panels (a) and (c) represent multi-vortex states.  Panels (b) and (d) are giant vortex states and (e) is a vortex-antivortex configuration.} \label{scheme1}
\end{figure}

In this section, we discuss the appearance and stability of anti-vortex states in the BdG theory in order to explain the existence of the giant antivortex.  Actually, such a state was already found in Ref. \cite{anti2} through the linear GL method by introducing artificial pinning.

From the phase diagram shown in Fig.~\ref{Phase}, we found that anti-vortex states are surprisingly stable within BdG theory.  This is due to the fact that the grid pattern oscillation of the order parameter gives an additional contribution to the symmetry of the vortex states and therefore, in a square sample, the $C_4$ symmetry is enhanced.  The other reason to form an anti-vortex is that the oscillations induced by the order parameter are seen in the vortex core where the order parameter is already suppressed.  These oscillations can easily lead to a shift in the phase of the order parameter by $\pi$ and, thus, result in the formation of vortex-antivortex molecules.

In order to explain this, we first discuss briefly the vortex profile in the GL theory. Fig.~\ref{scheme1} shows a schematic diagram of the vortex states for different winding number $L$ in GL theory.  The diagonal profiles of the order parameters vary smoothly in space and the vortex emerges where the order parameter vanishes.  Note that the phase of the order parameter is adjusted such that along the diagonal the order parameter is real. Panel (a) from Fig.~\ref{scheme1} shows the simplest case when only one vortex sits at the center.  As can been seen from the figure, the order parameter changes sign, which indicates the $\pi$ phase shift of the order parameter.  The profile is an odd function and $\Delta(r)\sim r$ near the vortex core.  Panels (b) and (c) from Fig.~\ref{scheme1} show the diagonal profiles for $L=2$.  Both profiles are even functions due to the $2\pi$ phase shift between the opposite corners.  The order parameter exhibits $\Delta(r)\sim r^2$ property.  When there is only one root, as can be seen from panel (b), the vortex is a giant one.  When there are two roots, as shown in panel (c), the configurations are multi-vortex states.  Similarly, the profiles of the order parameter shown in panels (d) and (e) from Fig.~\ref{scheme1} for $L=3$ show a $\Delta(r)\sim r^3$ spatial dependence.  One root means that we have a giant vortex state whereas three roots represent a vortex-antivortex configuration.  Note that, in order to generate the central anti-vortex, the order parameter has to oscillate around the center of the square.

\begin{figure}[t]
\includegraphics[width=\columnwidth]{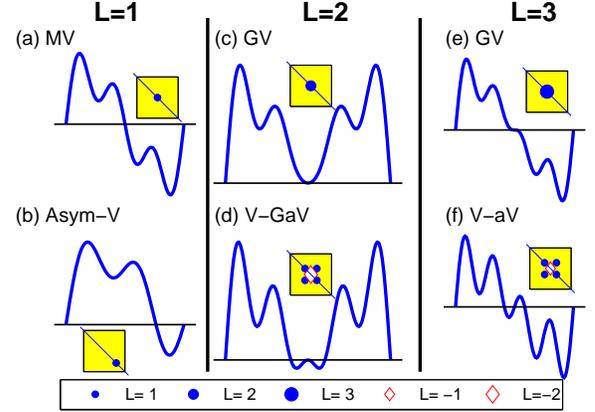}
\caption{(Color online) Similar as Fig.~\ref{scheme1} but now for the BdG theory.  Panels (a) and (b) show the symmetric and asymmetric $L=1$ vortex states, respectively.  Panels (c) and (d) are the giant vortex and vortex-giant antivortex $L=2$ states, respectively.  Panels (e) and (f) are the giant vortex and vortex-antivortex $L=3$ configurations, respectively.} \label{scheme2}
\end{figure}

\begin{figure}[ttt]
\includegraphics[width=\columnwidth]{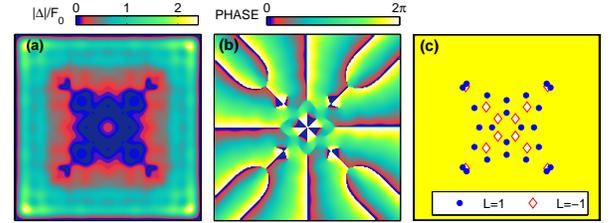}
\caption{(Color online) Vortex state for sample III with $W/\xi_0=7$ at $\Phi/\Phi_0=20$ with winding number $L=12$.  Panels (a) and (b) show contour plots of the order parameter and its phase, respectively.  Panel (c) shows schematically the vortex configuration.  Dark(blue) dots and open diamonds indicate vortices and antivortices, respectively.} \label{scheme3}
\end{figure}

Now let us move to nano-size superconductors where the BdG theory has to be used and the spatial oscillation of the order parameter cannot be neglected.  As can be seen from Fig.~\ref{scheme2} the oscillation plays an important role in generating vortices, especially when the value of the order parameter is comparable to the amplitude of the oscillation.  For instance, panels (a) and (b) from Fig.~\ref{scheme2} show the symmetric and the asymmetric $L=1$ vortex states.  The reason for the appearance of the asymmetric vortex state is the fact that the order parameter has an odd number of oscillations across the diagonal.  Thus, the vortex can not sit at the center.  For the $L=2$ states, panel (c) shows a giant vortex state where the sign of the profile of the order parameter is always positive across the diagonal.  However, due to the oscillations, the case shown in panel (d) of Fig.~\ref{scheme2} can easily exist and shows a giant anti-vortex ($L=-2$) at the center.  Further, the configurations show a large diversity for a fixed winding number $L$.  When $L=2$, the configuration can be $1+1$, $2+0$, $3-1$, $4-2$ and so on.  Panels (e) and (f) from Fig.~\ref{scheme2} are for $L=3$ states.  Apparently, they are similar to the GL case shown in Fig.~\ref{scheme1}, but the probability of the occurrence of the V-aV state is much larger than in GL case.  The reason is that the result with one root is just a special case while the general case shows oscillations at and around the vortex core.

The V-aV molecules do not only exist for smaller winding number $L$, but they can also appear for large $L$ in the BdG results.  Fig.~\ref{scheme3} shows an example for sample III with $W/\xi_0$=7 at $\Phi/\Phi_0=20$ and with a winding number $L=12$.  We find that vortices concentrate in the central dark (blue) area where the order parameter is strongly suppressed.  From the phase of the order parameter, which is shown in Fig.~\ref{scheme3}(b), the total winding number $L=12$ is found but it is difficult to distinguish each vortex.  After a careful analysis, we plot schematic diagram of the vortex configuration in Fig.~\ref{scheme3}(c).  The dark(blue) dots and open diamonds indicate vortices and anti-vortices, respectively.  As can be seen, the lattices of vortices and antivortices are nested within each other.  Since anti-vortices attract vortices, all the vortices ($24$ vortices and $12$ antivortices) can be condensed in the cental area of the sample.  This picture becomes more accurate when $k_F\xi_0$ is large.   The stronger the oscillations of the order parameter the more V-aV pairs are generated.  However, the size of the V-aV pair can only be of the order of the Fermi wavelength.  Thus, it will be very hard to detect them in experiments.  This is why these states are mostly treated as a giant vortex in conventional superconductors.  In other words, the suppressed central area of the order parameter, after coarse graining, will look like a giant vortex with $L=12$.

\section{Conclusion}

To summarize, we investigated the vortex states in a nanoscale superconducting square for different sizes $W$, parameters $k_F\xi_0$,  and temperatures $T$.  First, we found that the inhomogeneous pattern of the order parameter in the absence of magnetic field strongly depends on $k_F\xi_0$ and the size $W$.  This oscillation pattern will give an additional contribution  to competing effects that determine the vortex configurations when the field is applied. Due to the inhomogeneous order parameter induced by the quantum topological confinement, samples with different $k_F\xi_0$ and $W$ will favor different winding numbers $L$.

We find unconventional vortex states such as asymmetric, edge-parallel and vortex-antivortex states as the ground state of our nanoscale system. These were never seen in the Ginzburg-Landau approach. The inhomogeneous pattern of the order parameter, especially the strong oscillation at the boundaries causes additional potential wells for vortices which in turn generates a lot of metastable vortex states. Furthermore, in the quantum limit,  nano-size superconductors favor vortex-antivortex molecules while disfavoring giant vortex states.

We observe that vortex ground states and the phase transition fields are very sensitive to changes in the parameter $k_F\xi_0$, size $W$ and temperature $T$.  This is a direct consequence of the quantum size effect.  However, this effect is suppressed when the size $W$ is large or when temperature is high. In this case most metastable states become unstable and the ground states become compatible with GL theory.

For high magnetic fields, vortex-antivortex pairs can be easily found when $k_F\xi_0$ is large because the absolute value of the order parameter becomes smaller than the amplitude of its oscillations.  However, detection of such states is beyond the current experimental abilities.

The peculiar vortex states uncovered in the present work should be observable in superconducting systems where $k_F \xi_0$ is small. Such systems could be high-$T_c$ superconducting nano-grains for which the coherence length is small or cold-atom condensates with small $k_F$, i.e. large Fermi wavelength. Of special interest could be hybrid systems made of superconducting substrates and graphene sheets for which the Fermi wavelength is highly tunable near the Dirac point. Future work could also address the fundamental vortex-vortex and vortex-antivortex interactions for systems with a small  $k_F \xi_0$, for which the oscillations of the order parameter on the order of $\lambda_F$ become important.

\section{Acknowledgments}
This work was supported by the Flemish Science Foundation (FWO-Vlaanderen) and Methusalem Funding of the Flemish government.

\pagebreak

\end{document}